\documentclass[reprint,letterpaper,pre,showpacs]{revtex4-1}
\usepackage{graphicx}
\usepackage{times}

\newcommand{\kB}{\ensuremath{k_\mathrm{B}}}

\begin{document}

\title{Optimal reconstruction of the folding landscape 
using differential
energy surface analysis}

\author{Arthur \surname{La Porta}}
\author{Natalia A.\ Denesyuk}
\author{Michel \surname{de Messieres}}
\affiliation{University of Maryland College Park, Physics Department, Biophysics Program, Institute for Physical Sciences and Technology}
\date{\today}

\begin{abstract}In experiments and in simulations, the free energy 
of a state of a system can be determined from the probability that
the state is occupied.
However, it is often necessary to impose a biasing potential on the
system so that high energy states are sampled with sufficient
frequency.  
The unbiased energy is typically obtained from the data
using the weighted histogram
analysis method (WHAM).
Here we present differential energy surface analysis (DESA),
in which the gradient of the energy surface, $dE/dx$, is extracted from
data taken with a series of harmonic biasing potentials.
It is shown that
DESA produces a maximum likelihood 
estimate of the folding landscape gradient.  
DESA is demonstrated by analyzing data from a simulated system as well as 
data from a single-molecule unfolding experiment in which the end-to-end
distance of a DNA hairpin is measured.
It is shown that the energy surface obtained from DESA is indistinguishable
from the energy surface obtained when WHAM is applied to the same data.
Two criteria are 
defined which indicate whether the DESA results are self-consistent.
It is found that these criteria can detect a situation
where the energy is not a single-valued function of the measured
reaction coordinate.
The criteria were found to be satisfied for the experimental data
analyzed, confirming that end-to-end distance is a good reaction coordinate
for the experimental system.  The combination of
DESA and the optical trap assay in which a structure is disrupted under
harmonic constraint facilitates an extremely accurate
measurement of the folding energy surface.
\end{abstract}


\maketitle

\section{Introduction}

In a dynamical system driven by thermal fluctuations
the effective energy $E$ as a function of conformation
$\mathbf{x}$ is related to the probability $p$ 
that the conformation
is observed by the Boltzmann formula,
\begin{equation}
p(\mathbf{x}) = \exp\left(\frac{-E(\mathbf{x})}{\kB T}\right),
\label{eq:boltzmann}
\end{equation}
where $\kB$ is the Boltzmann constant and $T$ is the temperature.
The conformation of a simple system may be specified
by a small number of variables.  However, 
in studies of the folding of bio-polymers the
conformational space of the system has
many degrees of freedom. 
In some cases, such systems can
be described in terms of a single reaction coordinate, $x$,
and the dynamics of the system can be modeled by diffusion
in this 1D space under the influence of an effective
energy\cite{plotkin:2002,onuchic:2004,wolynes:1995,dill:2008}.
In numerical simulations the reaction coordinate 
may be the radius of gyration of the
structure, the fraction of native contacts, or another
measure of the level of compaction or organization
of the molecule.  In single molecule manipulation experiments
the end-to-end extension of the molecule is
typically used as a reaction coordinate\cite{liphardt:2001}.  
The energy as a function of the reaction coordinate 
follows from Eq.~\ref{eq:boltzmann} as
\begin{equation}
E(x) = -\kB T \ln (p(x)) + c.
\label{eq:energy}
\end{equation}
The arbitrary constant $c$ is included because the
energy of a system is only defined up to a additive constant.
Although this formula can be used, in principle, to determine
the energy surface from the probability density function,
this is only practical when the energy varies in a range which 
is narrow compared with $\kB T$.  
The exponential dependence of the probability density on $E$ means
that states with relative energy that is large compared with $\kB T$ will
be impossible to sample in a finite time.

One solution to this problem is to apply an external 
force field to the system which tends to bias it 
towards the regions of the reaction coordinate that would
otherwise be poorly sampled.
Often, this takes the form of a harmonic constraint, which adds
an additional term $\alpha(x-x_0)^2/2$ to the
energy, where $\alpha$ is the effective stiffness 
and $x_0$ is the origin of the constraint.
By selecting an appropriate value of $\alpha$ and varying
$x_0$, the system can be forced to visit various regions
of the reaction coordinate, allowing more uniform convergence
of statistics.  This technique, often referred to 
as umbrella sampling\cite{torrie:1974},
is widely used in simulations\cite{frenkel:2002}, and has been applied to single molecule
experiments\cite{demessieres:2011}.

We can still apply Eq.~\ref{eq:energy} to the system
with a specific configuration of the harmonic constraint,
but we will obtain a biased energy which is the sum
of the intrinsic energy and the energy of the constraint.  To
find the unbiased energy, we subtract the known
constraint energy, and obtain
\begin{equation}
E_j(x) = -\kB T \ln (p_j(x)) -\frac{1}{2} \alpha (x-x_j)^2 + c_j.
\label{eq:constrainedenergy}
\end{equation}
For each position of the constraint $x_j$ we obtain
a measurement of the energy surface $E_j(x)$ over the
region visited by the system.  Each local energy surface
$E_j(x)$ contains an independent constant $c_j$.  

If we wish to find the global energy surface,
defined over the entire
domain of $x$, we need to choose the constants $c_j$ and combine the local energy
landscapes $E_j(x)$ in a self-consistent manner.  
If there is substantial overlap between the domains of 
the local landscapes, the constants $c_j$ can be determined
by requiring that the energy surfaces corresponding to different
constraints are consistent in the overlap regions.

The weighted histogram analysis method (WHAM) has been formulated to reconstruct
the energy surface $E(x)$ from Monte Carlo or molecular dynamics simulations with
arbitrary biasing potentials\cite{kramers:1940,ferrenberg:1989,kumar:1992,boczko:1995}. The method provides an optimal estimate for the unbiased 
probability density $p(x)$, 
\begin{equation}
p(x) = \sum_i p_i(x) w_i(x),
\label{eq:WHAMdef}
\end{equation}
where $p_i(x)$ is the probability density sampled in biased simulation 
$i$ and the summation is over all simulations. 
In the case of a harmonic constraint centered at 
$x_i$, the weights $w_i(x)$ are given by
\begin{equation}
w_i(x) = \frac{M_i}{\displaystyle \sum_j M_j\exp\left[f_j-\frac{\alpha (x-x_j)^2}{2\kB T}\right]},
\label{eq:WHAM1}
\end{equation}
where $M_i$ is the total number of measurements in 
simulation $i$. The constants $f_i$ are defined implicitly by a system 
of nonlinear equations,
\begin{eqnarray}
\exp\left(-f_i\right) = \int dx \exp\left(-\frac{\alpha (x-x_i)^2}{2\kB T}\right)\nonumber\\
\times\frac{\displaystyle \sum_j\frac{H_j(x)}{dx}}{\displaystyle \sum_k M_k\exp\left[f_k-\frac{\alpha (x-x_k)^2}{2\kB T}\right]},
\label{eq:WHAM2}
\end{eqnarray}
where the histogram count $H_j(x)$ is the number 
of measurements between $x$ and $x+dx$ in system $j$,
and is related to the probability by $p_j(x) dx = H_j(x)/M_j$.

In the following section,
we describe another method of obtaining the global
energy surface which we call differential energy surface 
analysis (DESA).
In DESA, we consider the slope of the energy landscape,
$dE/dx$ rather than the energy itself.
Differentiating Eq.~\ref{eq:constrainedenergy} with
respect to $x$ we obtain
\begin{equation}
\frac{dE_j}{dx}(x) = -\kB T \frac{d}{dx}\left[\ln (p_j(x))\right] - \alpha (x-x_j).
\label{eq:constraineddE}
\end{equation}
An important feature of this equation is that
the constants $c_j$ are eliminated, so that it is not
necessary to find a self-consistent solution to 
obtain the global function $dE/dx$.  
At any given point $x$ along the landscape, $dE/dx$
can be obtained by averaging $dE_j/dx$ obtained from
the system at various constraint origins.

\section{Description of DESA}

In order to define the method of differential
energy landscape analysis, we assume a thermally
driven system with one reaction coordinate $x$ which is characterized
by an energy function $E(x)$.  We assume that
the dynamics of the system are measured in the presence of
a harmonic constraint of stiffness $\alpha$ for $N$ distinct 
constraint origins $x_j$.  
For each $x_j$, the time series of $x$ is used
to compile a histogram $H_j(x_i)$   
containing $M_j$ total samples.   We  
assume that the histogram binning is consistent for all $x_j$,
and that the values of $\alpha$ and 
$x_j$ are chosen so that there is significant overlap between
the histograms.  The slope of the energy landscape
$dE/dx$ at position $x_i$ is given by
\begin{equation}
\frac{dE}{dx}(x_i) =%
\frac{\sum_j H_j(x_i) \frac{dE_j}{dx}(x_i)}{\sum_j H_j(x_i)},
\label{eq:DESAdef}
\end{equation}
where the summation is over the constraint origins, $x_j$,  
and $\frac{dE_j}{dx}(x_i)$ is defined by Eq.~\ref{eq:constraineddE}.
Interpreting this formula, the value of $dE/dx$ at position
$x_i$ is a weighted average of $dE_j/dx$ found
from the $N$ systems with constraint origins $x_j$.
Using $p_j(x_i) \Delta x = H_j(x_i)/M_j$, we can
express Eq.~\ref{eq:DESAdef} entirely in terms of histogram
counts, as
\begin{eqnarray}
\lefteqn{\frac{dE}{dx}(x_i)=}&&\nonumber\\
&&\frac{\sum_j \left[-\kB T \frac{d}{dx}\left(\ln H_j(x_i)\right)-\alpha(x_i-x_j)\right]H_j(x_i)}{\sum_j H_j(x_i)}.
\label{eq:DESArewrite}
\end{eqnarray}
This formula has been used to reconstruct energy landscapes of molecular
dynamics simulations\cite{denesyuk:2009}, and experimental data\cite{demessieres:2011}.
We will show below that Eq.~\ref{eq:DESArewrite}
gives an optimal estimation of $dE/dx$.

When determining the mean value of a Gaussian distributed variable 
from uncorrelated data points
which have differing uncertainty, the maximum likelihood
solution is
\begin{equation}
\bar{a} = \frac{\sum_i w_i a_i}{\sum w_i},\quad \sigma_{\bar{a}}^2 = \frac{1}{\sum_i w_i},\quad  w_i = \frac{1}{\sigma_i^2},
\label{eq:maxL}
\end{equation}
where $\sigma_i$ is the standard deviation of the statistical
ensemble from which $a_i$ is taken, $\bar{a}$ is the mean of $a$
and $\sigma_{\bar{a}}$ is the standard deviation of $\bar{a}$.  In order to show that
Eq.~\ref{eq:DESAdef} is a maximum likelihood estimate
of $dE/dx$ we must show that the choice $w_j(x_i) = H_j(x_i)$
meets the criteria set out in Eq.~\ref{eq:maxL}.

Starting with Eq.~\ref{eq:constraineddE}, the evaluation of $dE/dx$
will involve a finite difference of the natural logarithm of $H_j(x_i)$,
\begin{eqnarray}
\frac{dE_j}{dx}(x_i) &=&\frac{-\kB T}{\Delta x}%
\left[ \ln\left(H_{j}(x_{i+1}) \pm \Delta H_{j}(x_{i+1})\right)\right.\nonumber\\
&&\left.-\ln\left(H_{j}(x_{i-1}) \pm \Delta H_{j}(x_{i-1})\right)\right],
\label{eq:DESAuncertain}
\end{eqnarray}
where $\Delta H_{j}(x_i)$ represents the statistical uncertainty in $H_{j}(x_i)$, 
and $\Delta x = x_{i+1} - x_{i-1}$.
Note that any terms with $H_j(x_{i})$ are suppressed in Eq.~\ref{eq:DESAdef},
but we also must suppress any terms where $H_j(x_{i-1})$ or $H_j(x_{i+1})$
is zero, since in this case the derivative is undefined.
We can re-express Eq.~\ref{eq:DESAuncertain} as
\begin{eqnarray}
\frac{dE_j}{dx}(x_i)&=&\frac{\kB T}{\Delta x}\left[ \ln \left(H_{j}(x_{i+1})\left(1 \pm \frac{\Delta H_{j}(x_{i+1})}{H_{j}(x_{i+1})}\right)\right)\right.\nonumber\\
&&- \left.\ln\left(H_{j}(x_{i-1})\left(1 \pm \frac{\Delta H_{j}(x_{i-1})}{H_{j}(x_{i-1})}\right)\right) \right]\nonumber\\
&=&\frac{\kB T}{\Delta x}\left[\rule{0pt}{15pt}\ln \left(H_{j}(x_{i+1})\right) - \ln\left(H_{j}(x_{i-1})\right) \right.\nonumber \\
&&\ \ + \ln\left(1 \pm \frac{\Delta H_{j}(x_{i+1})}{H_{j}(x_{i+1})}\right)\nonumber\\
&&- \left. \ln \left(1 \pm \frac{\Delta H_{j}(x_{i-1})}{H_{j}(x_{i-1})}\right)\right],
\label{eq:DESAuncertain2}
\end{eqnarray}
so that the uncertainties in $H_{j}(x_{i-1})$ and $H_{j}(x_{i+1})$ produce 
additive uncertainties in $dE_j/dx$.
Assuming the uncertainty in $H_j(x_i)$ is statistical, the uncertainty
terms can be simplified using $\Delta H_j(x_i) = \sqrt{H_j(x_i)}$,
so that
\begin{equation}
\ln\left(1 \pm \frac{\Delta H_{j}(x_{i})}{H_{j}(x_{i})}\right) =%
\ln\left(1 \pm \frac{\sqrt{H_{j}(x_{i})}}{H_{j}(x_{i})}\right) %
\approx \frac{\pm 1}{\sqrt{H_{j}(x_{i})}},
\label{eq:logsimplify}
\end{equation}
where the last step is an expansion of the expression to first order.
Since $\Delta H_j(x_{i-1})$ and $\Delta H_j(x_{i+1})$ are uncorrelated, the 
errors arising from these terms add in quadrature.  In the limit that $\Delta x$
is small compared with any important features of the energy landscape 
we can neglect the difference between $H_j(x_{i-1})$ and $H_j(x_{i+1})$,
and replace both by $H_j(x_{i})$.
Using Eq.~\ref{eq:logsimplify} we can then approximate the uncertainty in $\frac{dE_j}{dx}$
as
\begin{equation}
\sigma_j = \frac{\kB T \sqrt{2}}{\Delta x \sqrt{H_j(x_i)}}.
\label{eq:sigmaj}
\end{equation}
The statistical weight required for maximum likelihood is therefore
\begin{equation}
w_j(x_i) = \frac{1}{\sigma_j^2} = \left(\frac{\Delta x}{\kB T }\right)^2\frac{H_j(x_i)}{2}.
\end{equation}
Since an overall multiplicative factor will cancel out in
Eq.~\ref{eq:maxL} and not affect the calculation of the
mean value, Eq.~\ref{eq:DESAdef} is equivalent to the maximum
likelihood estimation of $dE/dx$ and is an optimal estimation.

\section{Diagnostics in the DESA method}
\label{sec:diagnostics}

\begin{figure}[tb]
\begin{center}\includegraphics[width=2.75in]{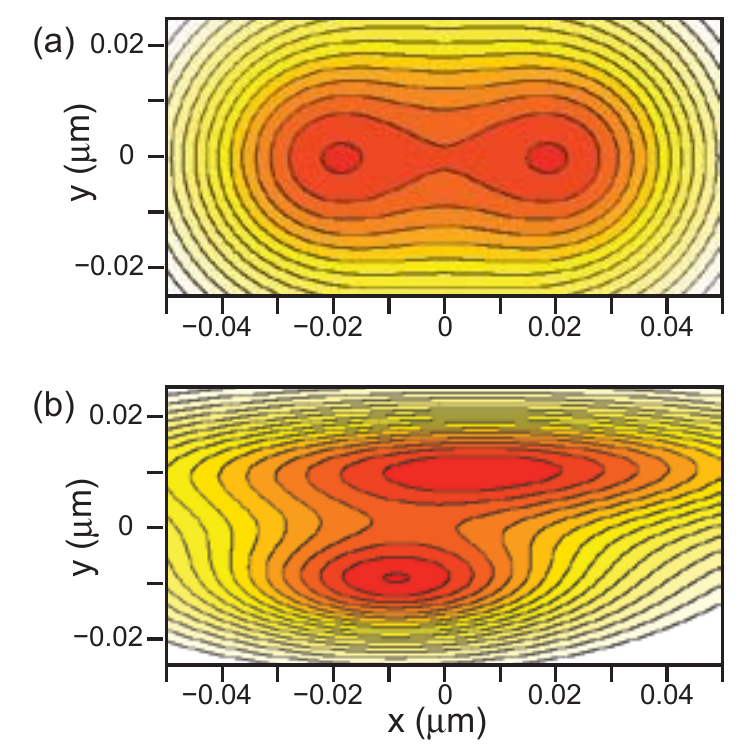}\end{center}
\caption{Contour maps of energy surfaces defined by Eq.~\ref{eq:Esurface}, where energy
is measured in pN$\cdot$$\mu$m and distance is measured in $\mu$m.  (a)  System I 
with $A_1=A_2=0.15$ and wells centered at ($-0.015$, $0$) and ($0.015$, $0$) with width ($0.02, $0.02).  
Contours are spaced by 0.0075. (b)  System II with $A_1=A_2=0.20$.  
The first well has center ($-0.009$, $-0.011$) with width
($0.04$, $0.0125$) and the second well has center ($0.009$, $0.011$) 
with width ($0.08$, $0.0125$).  Contours are spaced by 0.01. }
\label{fig:Mcontour}
\end{figure}

\begin{figure}[tb]
\begin{center}\includegraphics[width=2.75in]{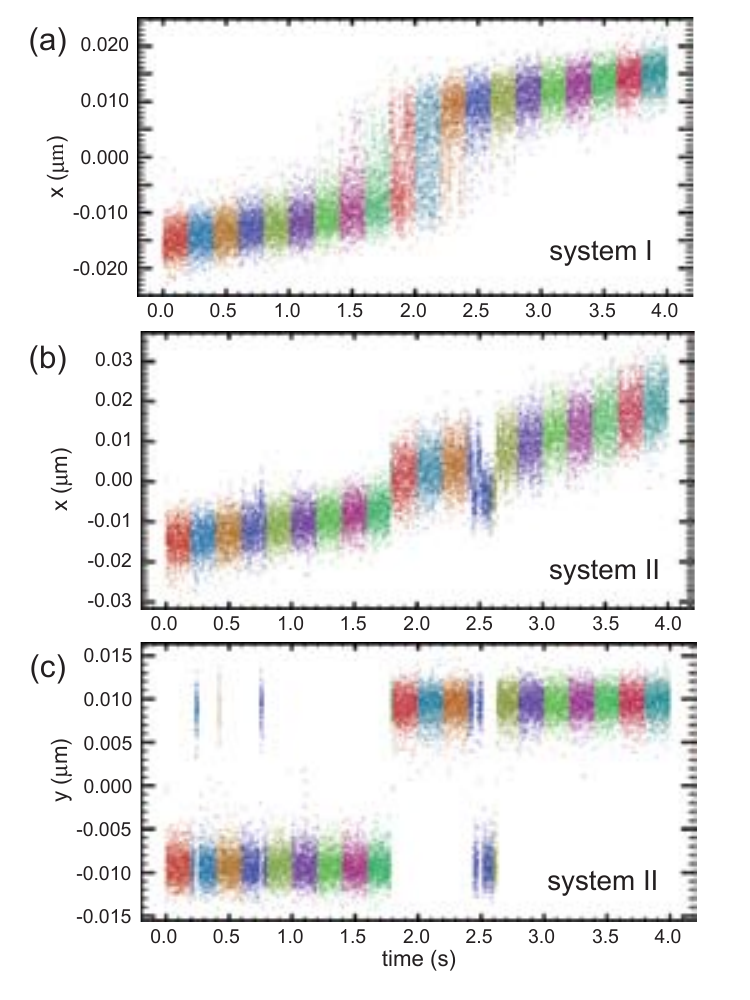}\end{center}
\caption{Trajectories for simulated systems. (a) $x$ coordinate for system I.  
The $y$ coordinate of system I fluctuations around zero (data not shown).  
(b)  $x$ coordinate for system II.
(c) $y$ coordinate for system II.}
\label{fig:Strajectory}
\end{figure}

\begin{figure*}[tb]
\begin{center}\includegraphics[width=6.5in]{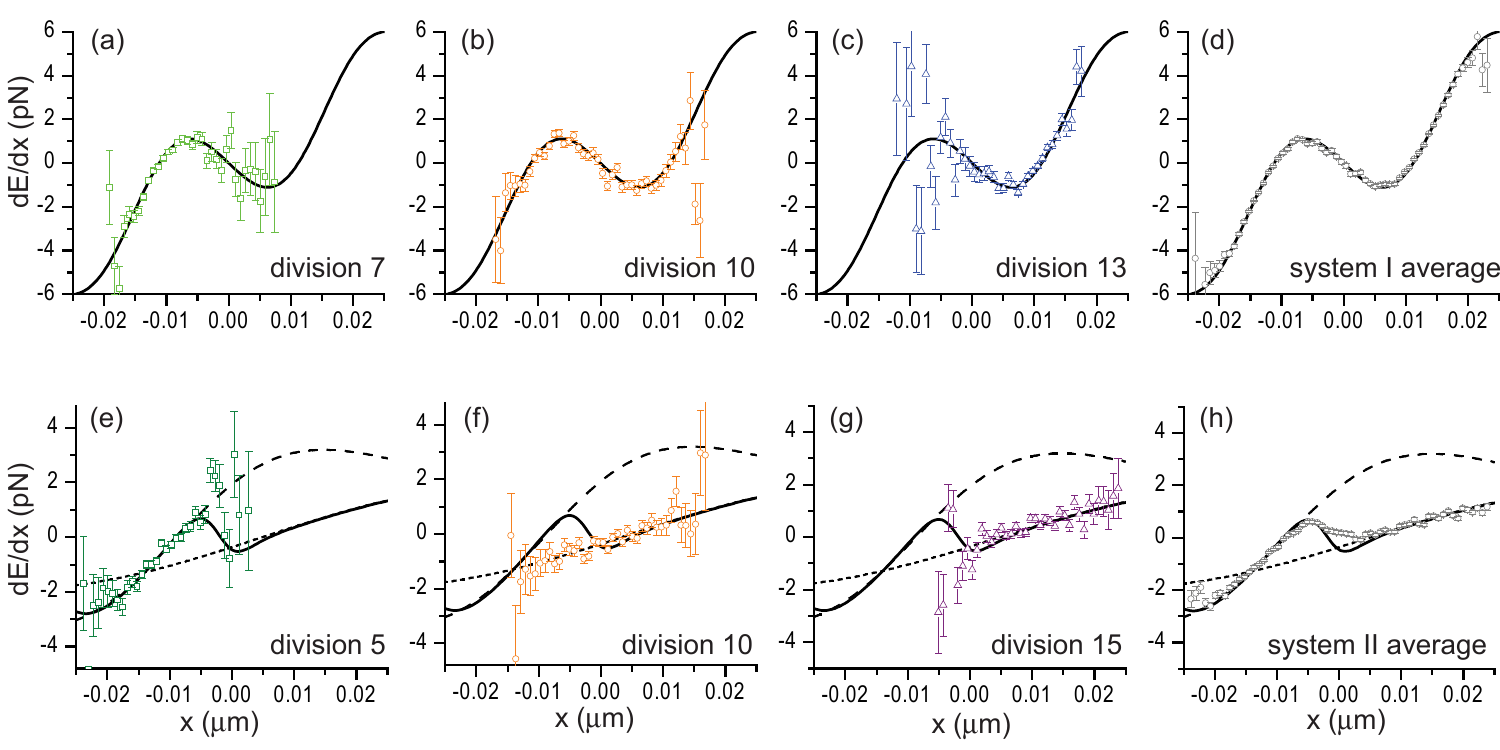}\end{center}
\caption{Reconstruction of the derivative of the energy landscape
for the two simulated systems.  In (a)-(c) $dE/dx$ is calculated 
by applying Eq.~\ref{eq:constraineddE} to
divisions 7, 10 and 13 of the data shown in Fig.~\ref{fig:Strajectory}(a)
and in (d) the reconstruction of $dE/dx$ based on Eq.~\ref{eq:DESArewrite} is
shown.
The solid curve is $dE/dx$ as a function of $x$ calculated from the simulation potential 
assuming that $y$ values are occupied with statistical weight proportional to the 
Boltzmann factor.
In (e)-(g) $dE/dx$ is shown for divisions 5, 10 and 15 of the data shown in
Fig.~\ref{fig:Strajectory}(b) and in (h) the reconstruction of $dE/dx$ is shown.
The solid curve is $dE/dx$ as a function of $x$ from the simulation potential 
assuming that $y$ values are occupied with statistical 
weight proportional to the 
Boltzmann factor, and the long and short dashed lines represent
$dE/dx$ but assuming the system is confined to negative or positive $y$, respectively.}
\label{fig:SdEdx}
\end{figure*}

\begin{figure}[tb]
\begin{center}\includegraphics[width=2.75in]{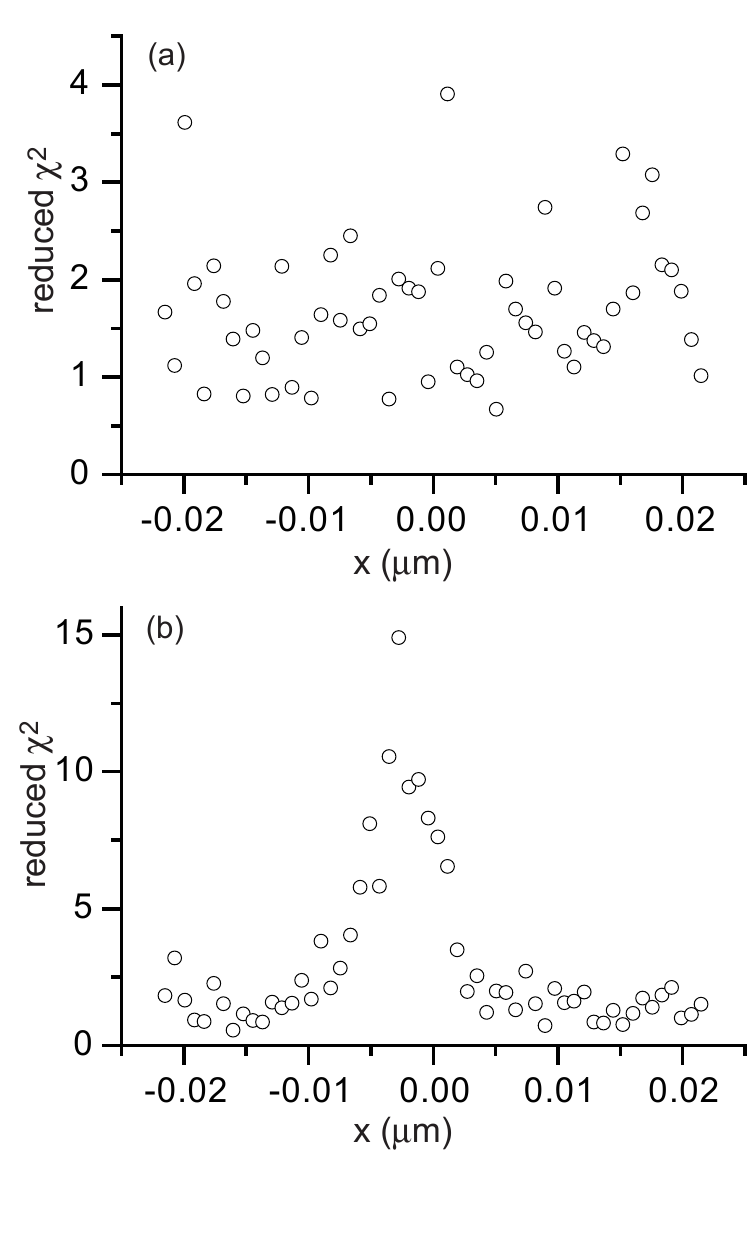}\end{center}
\caption{Reduced $\chi^2$ as a function of $x$ evaluated using
Eq.~\ref{eq:chi2} for system I (a) and system II (b).}
\label{fig:Schi2}
\end{figure}

\begin{figure}[tb]
\begin{center}\includegraphics[width=2.75in]{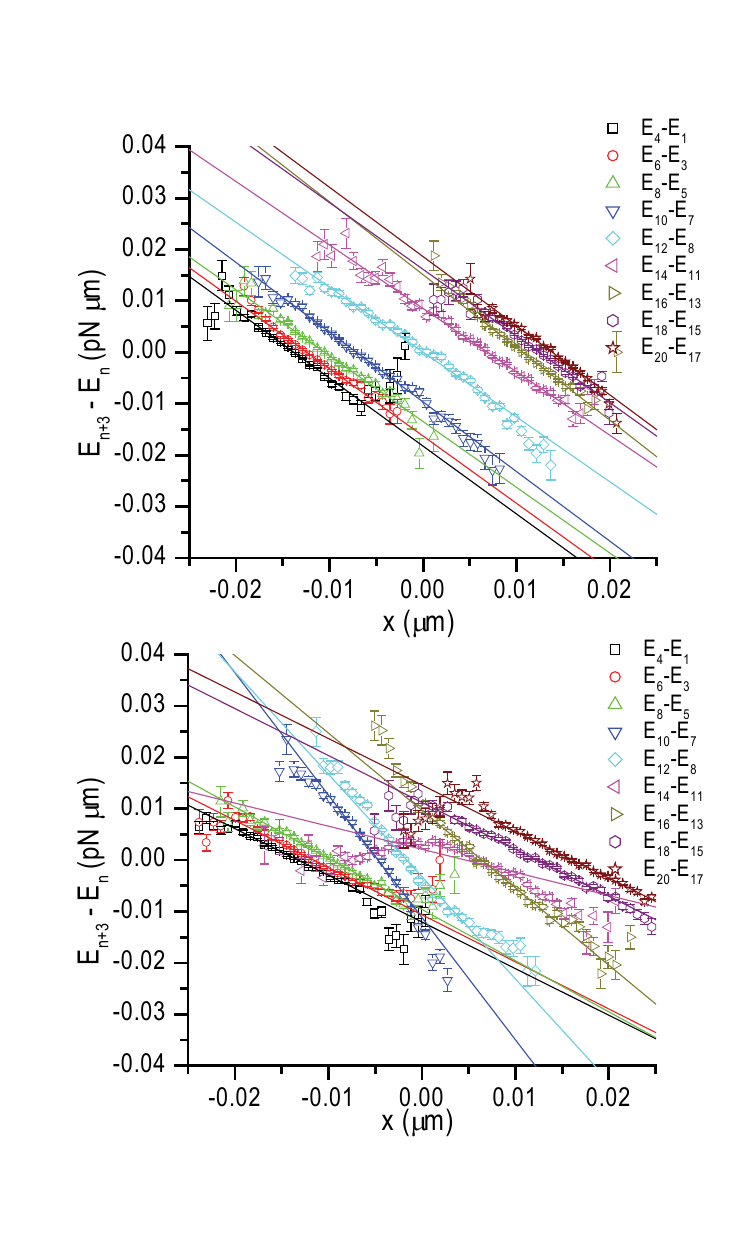}\end{center}
\caption{The energy difference defined by Eq.~\ref{eq:diag1} is plotted for
system I (a) and system II (b).  The energy
for each division $n$ is compared with the energy of division $n-3$.}
\label{fig:Sslope}
\end{figure}

\begin{figure}[tb]
\begin{center}\includegraphics[width=2.75in]{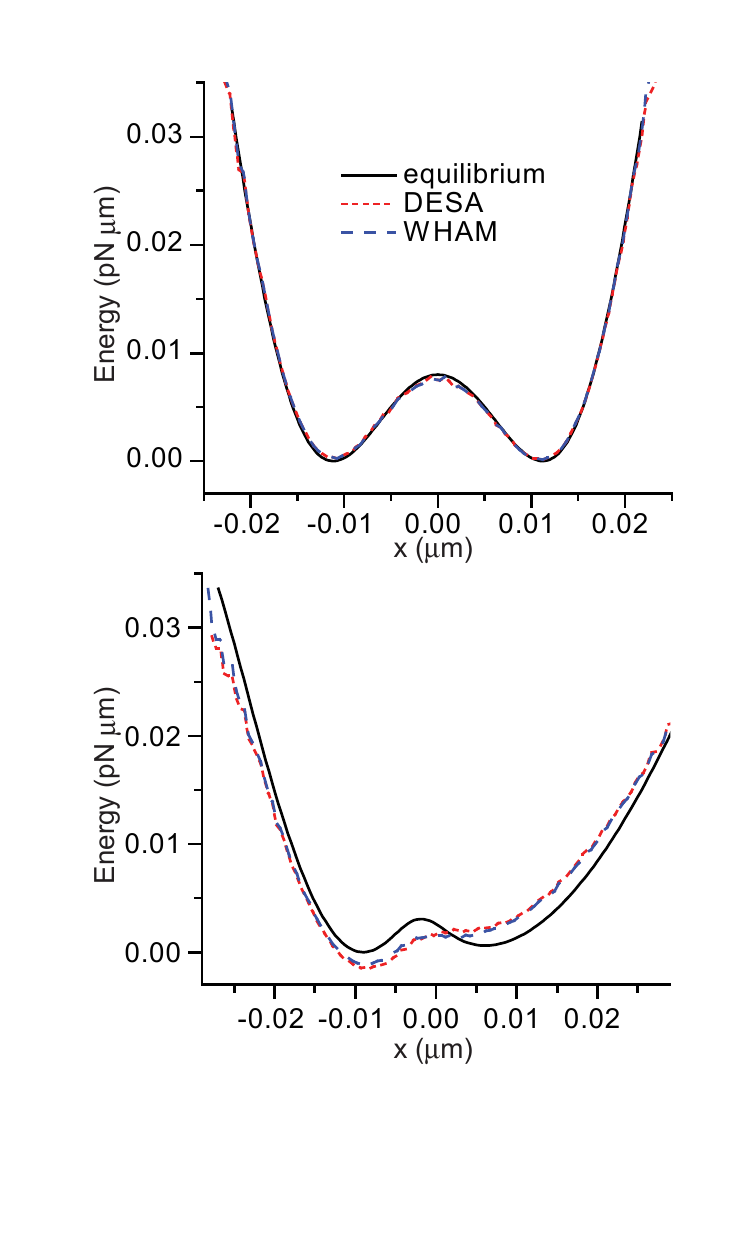}\end{center}
\caption{The energy surface obtained from integration of the DESA $dE/dx$ function 
is plotted using a short dashed line and the energy obtained by WHAM is plotted
using a long dashed line.  The solid line is the energy of the system as a function
of $x$, assuming that the system remains in thermal equilibrium with respect to $y$.
Data for system I is shown in (a) and data for system II is shown in (b).}
\label{fig:SEnergy}
\end{figure}

Recent work has provided criteria for error estimation in free energy 
calculations based on the weighted histogram analysis method\cite{zhu:2011,chodera:2007}.  
The DESA result is obtained by straightforward averaging of 
$dE_j/dx$ estimates obtained with different constraint origins $x_j$.
Use of the maximum likelihood estimation assures that the
optimal value of $dE/dx$ is produced, and straightforward error
propagation can be used to obtain the uncertainty in the
values of $dE/dx$ obtained. 
However, when employing umbrella sampling, it is necessary
to assume that the histograms obtained for different 
constraint origins overlap and that the data acquired with different constraint
origins are sampling the same energy surface.
One potential pitfall of umbrella sampling---whether WHAM or DESA is used
for analysis of the data---is that we 
can obtain a smooth measured energy surface even
if the energy is not a single-valued function of the reaction coordinate.
Here we introduce two criteria that can 
be applied in order to detect inconsistencies in $dE_j/dx$ values obtained 
from different constraint origins.  We will later show that these criteria give
a warning when the reconstructed landscape is not accurate.

The first method involves comparison of the biased energy surfaces
obtained from different constraint origins.  
Subtracting two biased energies, we obtain
\begin{eqnarray}
\lefteqn{E_{\mathrm{b},k}(x) - E_{\mathrm{b},j}(x) = } &&\nonumber\\
&&\left[E(x) + \frac{\alpha}{2}(x - x_k)^2\right]- \left[E(x) + \frac{\alpha}{2}(x - x_j)^2\right]\nonumber\\
&& = x \left[ \alpha(x_j - x_k)\right] + \frac{\alpha}{2}\left(x_k^2 - x_j^2\right),
\label{eq:diag1}
\end{eqnarray}
where $E_{\mathrm{b},j}(x)$ is the biased energy surface measured
with constraint origin $x_j$ and $E(x)$ is the
unbiased energy of the system.
The cancelation of $E(x)$ leaves 
terms which depend only on the biasing potential.
The constant term is not of interest, since the energy itself
is only defined up to an additive constant.  However, we expect
the energy difference to manifest a straight line with slope determined by the
constraint strength and the relative constraint displacement,
$\alpha(x_j - x_k)$.
If a different effective energy surface is in effect after the constraint
origin bas been moved, $E$ will fail to cancel in Eq.~\ref{eq:diag1}
and anomalous features will appear in the difference curve.

We can also test the self-consistency of the DESA analysis 
by determining if $dE_j/dx$ values obtained from individual constraint
origins deviate from the mean value in a manner that is consistent
with their statistical uncertainty.
For each histogram bin $i$ corresponding to 
position $x_i$, we evaluate
\begin{equation}
\chi^2 = \frac{1}{N-1}\sum_j\frac{\left( \frac{dE_j}{dx} - \frac{dE}{dx}\right)^2}{\sigma_j^2},
\label{eq:chi2}
\end{equation}
where $\sigma_j$ is the uncertainty in the value of $\frac{dE_j}{dx}$
obtained from the $j^{\mathrm{th}}$ constraint position (Eq.~\ref{eq:sigmaj}).
If the deviation of the individual values of $dE_j/dx$ from the mean are consistent
with the statistical uncertainty, the value of $\chi^2$ should be
of order 1. A value significantly larger than 1 indicates that systematic errors
are present in the $dE_j/dx$ values.

\section{Application of DESA to a simulated system}

\begin{figure}[tb]
\begin{center}\includegraphics[width=2.75in]{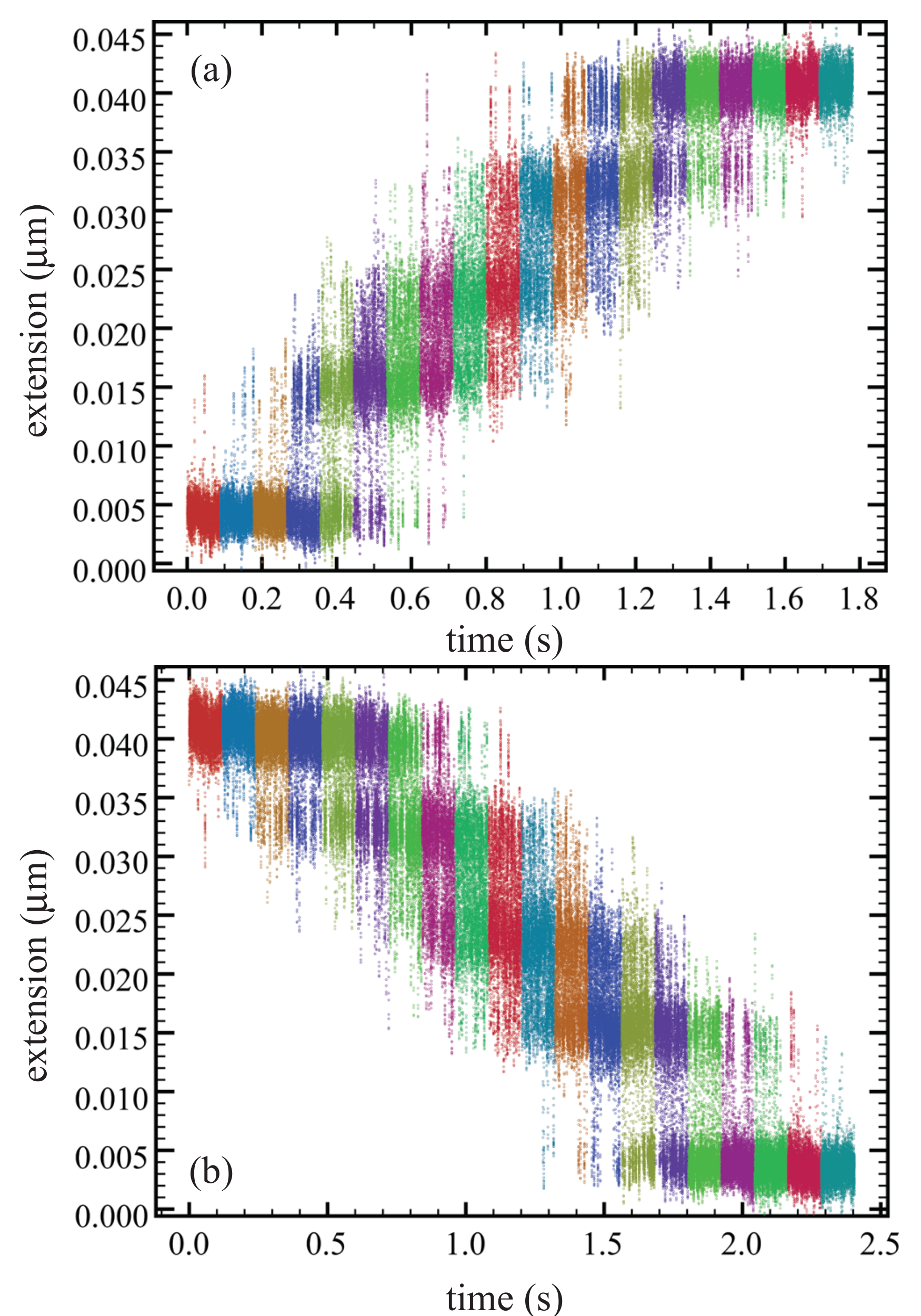}\end{center}
\caption{The extension
of a DNA hairpin as a function of time as the constraint origin is moved.
In (a) the hairpin is initially closed and the constraint moves at 25~nm/s.
In (b) the hairpin is initially open and the constraint moves at $-$25~nm/s.}
\label{fig:timeseries}
\end{figure}

\begin{figure*}[tb]
\begin{center}\includegraphics[width=6.0in]{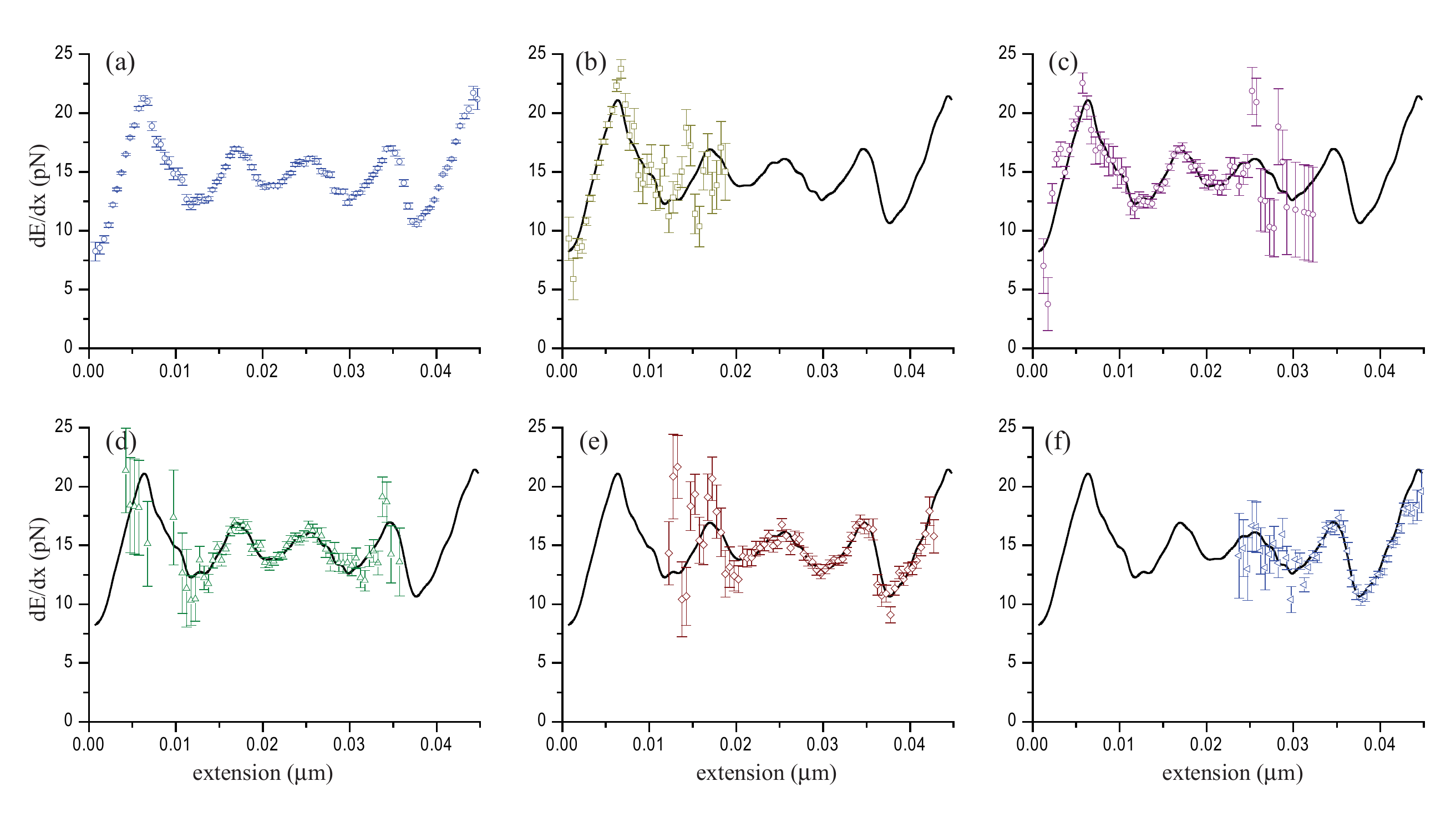}\end{center}
\caption{(a) Reconstructed $dE/dx$ plot.  In (b)-(f) the
$dE_j/dx$ estimates from the 3rd, 6th, 9th, 12th and 15th
intervals are compared with the reconstructed $dE/dx$
function.  The $dE_j/dx$ values are calculated using
Eq.~\ref{eq:constraineddE}.  For (a) the uncertainty is based on the maximum likelihood
result given in Eq.~\ref{eq:maxL} for (b)-(f) 
uncertainty is based on Eq.~\ref{eq:DESAuncertain2}.}
\label{fig:dEdxD}
\end{figure*}

\begin{figure}[tb]
\begin{center}\includegraphics[width=2.75in]{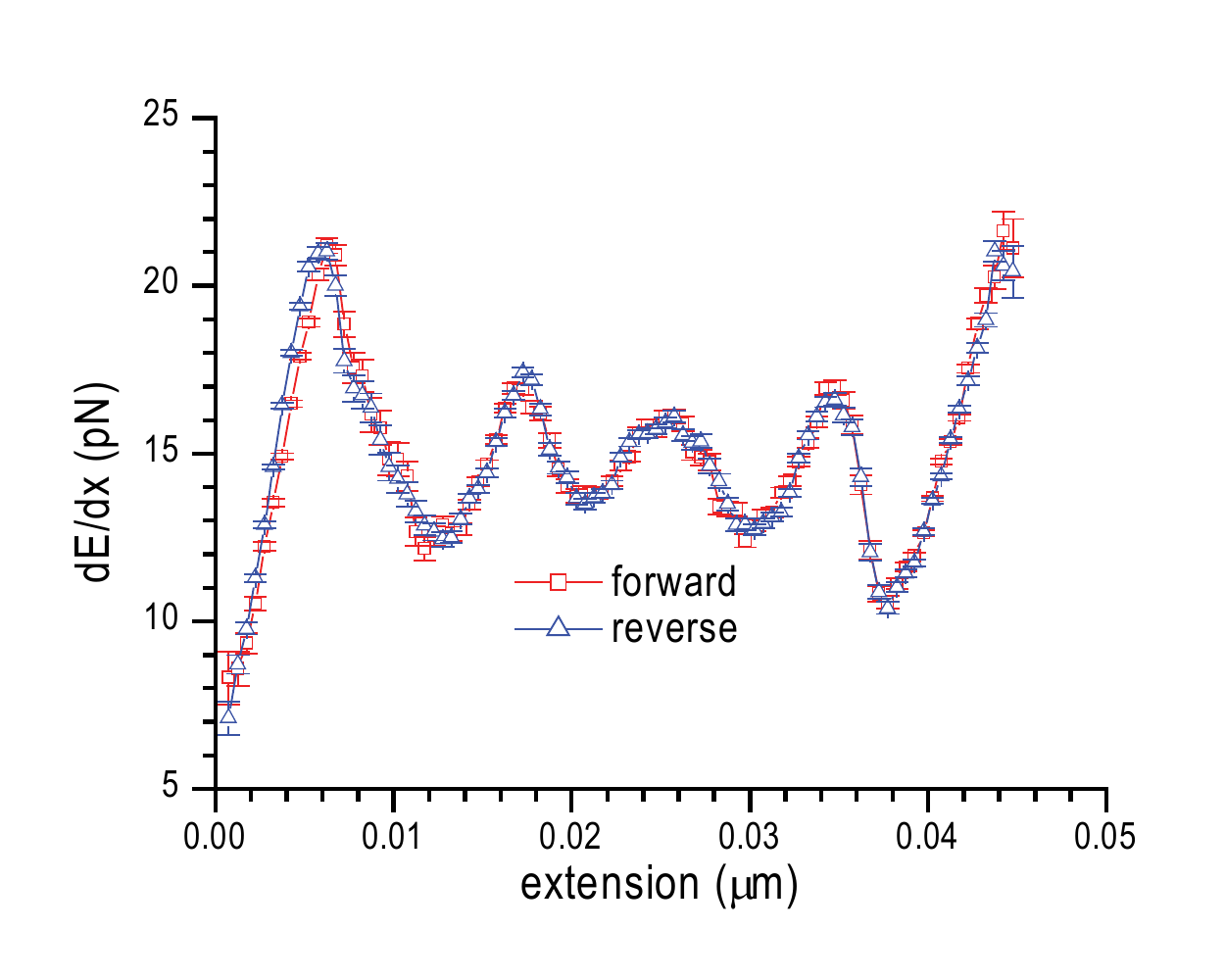}\end{center}
\caption{The $dE/dx$ curves obtained from unfolding and folding of the hairpin
are compared.}
\label{fig:reverse}
\end{figure}

\begin{figure}[tb]
\begin{center}\includegraphics[width=2.75in]{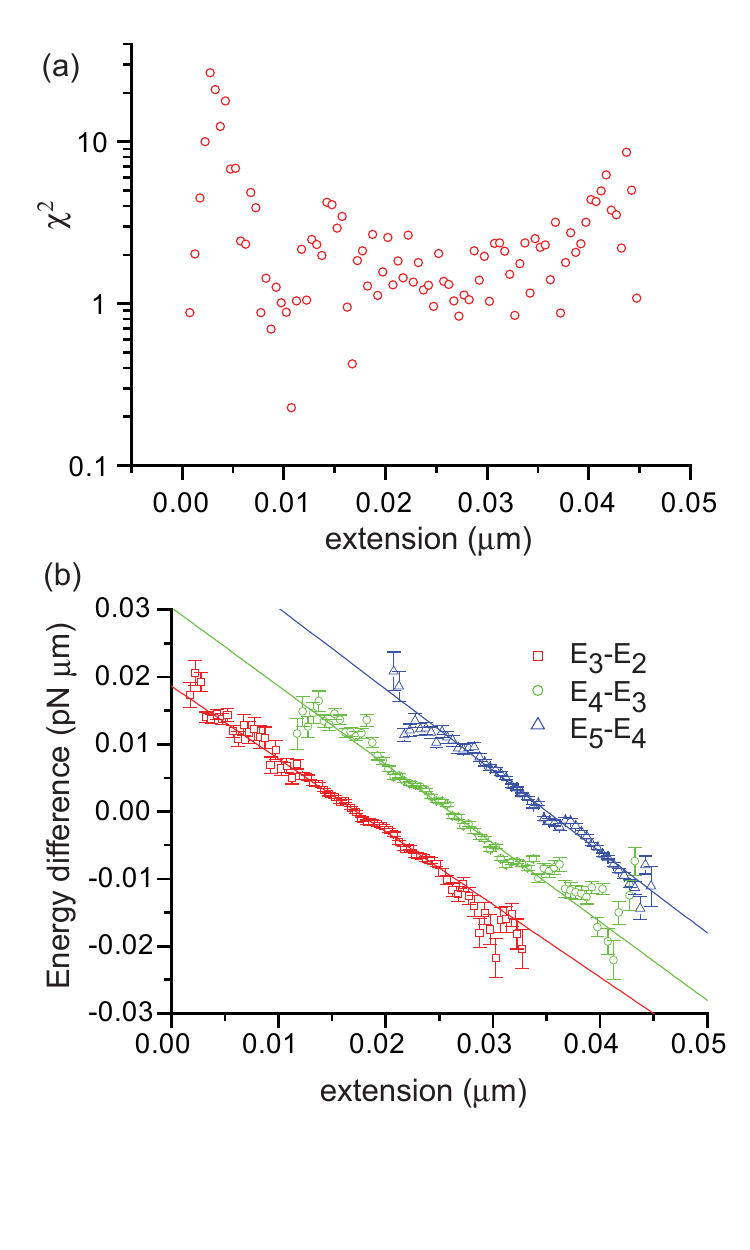}\end{center}
\caption{(a) The value of $\chi^2$ as a function of extension, for the
data shown in Fig.~\ref{fig:timeseries}(a).  (b) The difference in biased 
energy for adjacent intervals where the time series is divided
into six intervals rather than twenty.  The differences $E_3(x)-E_2(x)$, 
$E_4(x)-E_3(x)$ and $E_5(x)-E_4(x)$ are shown.}
\label{fig:Ediagnostic}
\end{figure}

\begin{figure}[tb]
\begin{center}\includegraphics[width=2.75in]{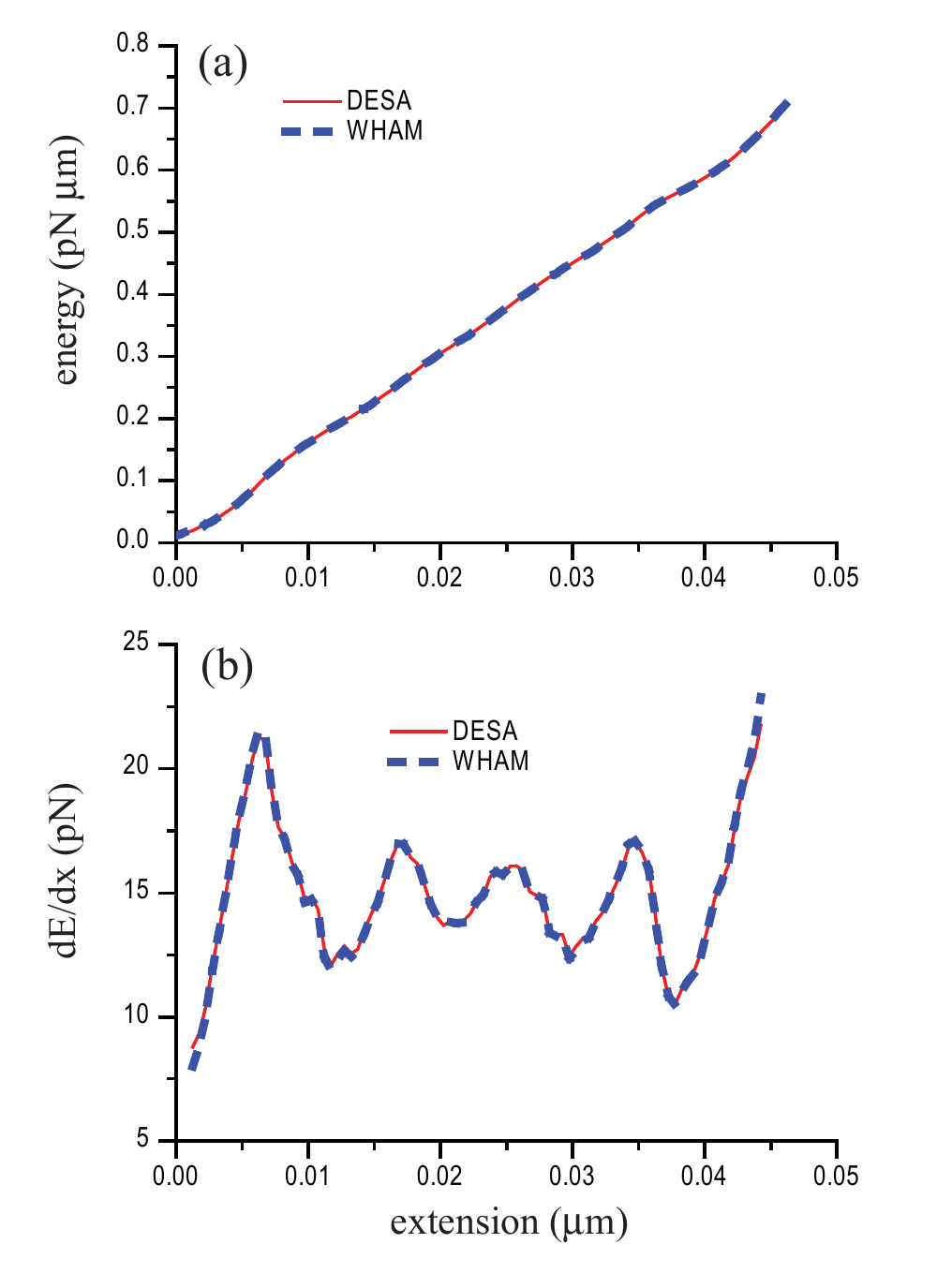}\end{center}
\caption{(a) Comparison of the energy surface obtained by DESA
and by WHAM from the time series in Fig.~\ref{fig:timeseries}(a).
(b) Comparison of $dE/dx$ obtained by DESA and WHAM from the time 
series in Fig.~\ref{fig:timeseries}(a).}
\label{fig:DESAWHAM}
\end{figure}

WHAM produces an optimal estimation of $p$ which is closely
related to the energy $E$ and DESA produces an optimal estimation of $dE/dx$.
For a well-behaved system with good statistical convergence we expect
both methods to converge to the underlying energy surface.
However, in simulations and in experiments it is often a challenge to obtain
adequate statistics, or obtain a reaction coordinate which unambiguously
specifies the state of the system.

In Section~\ref{sec:experiment} below we will apply DESA and WHAM to an experimental
system and compare the results.  
In this section we will apply DESA and WHAM to two simulated systems
in order to evaluate the accuracy with which the known energy
surface is obtained and illustrate the use of the diagnostic
criteria that were introduced in Section~\ref{sec:diagnostics}.
Both simulated systems involved diffusion on a 2D energy surface 
with two stable states and in both cases 
we assume that only one coordinate ($x$) is measured and that
the biasing potential is a function of $x$ only.  
Contour maps for the potential functions for the two systems are shown
in Fig.~\ref{fig:Mcontour}.
For both cases, the landscape consists of two overlapping 
potential wells with 2-dimensional Lorentzian profile.  
The form of the potential is 
\begin{eqnarray}
E(x,y)&=&\frac{-A_1}{1 + \frac{(x-x_1)^2}{{s_{x,1}}^2} + \frac{(y-y_1)^2}{{s_{y,1}}^2}}\nonumber\\
&&+\frac{-A_2}{1 + \frac{(x-x_2)^2}{{s_{x,2}}^2} + \frac{(y-y_2)^2}{{s_{y,2}}^2}}
\label{eq:Esurface}
\end{eqnarray}
where $A_n$ specifies the depth of each well, ($x_n$, $y_n$) specifies the center
and  ($s_{x,n}$, $s_{y,n}$) specifies the width
of each well in the $x$ and $y$ direction.

In system I, illustrated by Fig.~\ref{fig:Mcontour}(a),
there are two symmetrical potential wells lying on the $x$
axis (parameters given in the Fig.~\ref{fig:Mcontour} caption).
For this potential there is only one stable value of $y$ for each $x$.
In system II,  illustrated by Fig.~\ref{fig:Mcontour}(b)
the two potential wells have different width and are displaced in $y$
as well as $x$.
In system II
there is more than one stable value of $y$ for a given
value of $x$ and the measurement of $x$ is not sufficient
to determine the state of the system.   
The  transition between the two stable states of the system
involves a change in the unmeasured variable $y$.
Both simulated systems could serve as a model for a 
single-molecule unfolding experiment 
(such as the one described in Section~\ref{sec:experiment}) 
where a quantity
such as the end-to-end distance of the structure is under
experimental control but other undetectable degrees
of freedom are present.
In system I, the measured variable is a good reaction coordinate and
in the system II it is not.

The energy surfaces are used as the basis of a strongly damped Langevin simulation
with thermal energy $k_B T = 4.11 \times 10^{-4}\ \mathrm{pN}\cdot\mu\mathrm{m}$ and
drag coefficient $0.05\ \mathrm{pN} \cdot \mathrm{s}/\mu\mathrm{m}$.  In the simulation
400 time steps of $5 \times 10^{-8}\ \mathrm{s}$ 
were taken between each tabulated sample point. 
These parameters were chosen so that the energy and time scales of the simulated systems
roughly correspond to those
of the experimental system which we describe in Section~\ref{sec:experiment}.  
As a result, simulated and experimental runs of equal time
result in comparable statistical sampling.
Both simulation systems are run with a harmonic biasing potential 
which is continuously swept from
negative $x$ to positive $x$ to sample the transition. 
In system I the constraint stiffness is 150~pN/nm 
and the constraint origin sweeps from -0.03~$\mu$m to 0.03~$\mu$m over 4~s
and in system II the stiffness is 100~pN/nm and the 
origin sweeps the same range of position.

The trajectories obtained for the two versions of the simulation are shown in
Fig.~\ref{fig:Strajectory}.  In system I, the biasing potential causes the system
to be swept through the transition state with good sampling over the domain of the
reaction coordinate $x$.  
(In the course of the transition, $y$ fluctuates around zero, data not shown.)  In system II,
the biasing potential also produces relatively uniform sampling of $x$, although $y$
makes several abrupt transition between the basins of attraction at (-0.009~$\mu$m,-0.011~$\mu$m)
and (0.009~$\mu$m,0.011~$\mu$m).  The potential well at 
positive $y$ is more extended in $x$ than the one at negative $y$, 
resulting in larger fluctuations
in $x$ when $y$ is positive. 

The record of $x$ vs.\ time of the trajectories is divided into 20
equal time intervals and the histogram
of position is calculated for each interval.  Data for
each division is analyzed
using the constant constraint stiffness and the average position of
the constraint origin.  
In this simulation, it would be more natural to move the constraint origin in discrete
steps and hold it constant as each histogram is collected.
We move it continuously to more closely model the experimental procedure
used in the experiment described in Section~\ref{sec:experiment}.
In order to apply DESA or WHAM the constraint must be moved sufficiently slowly that
the system remains in quasi-equilibrium with respect to $x$ 
as the constraint origin moves.
We have chosen the simulation parameters 
to ensure that this condition is satisfied for both systems.

In Fig.~\ref{fig:SdEdx} the reconstruction of $dE/dx$ from the simulated 
systems is shown.
In Fig.~\ref{fig:SdEdx}(a)-(c) Eq.~\ref{eq:constraineddE} is used to obtain 
an estimation of
$dE_j/dx$ from three representative divisions of the system I 
trajectory.  The
three curves cover overlapping ranges of the reaction coordinate $x$.
The $dE_j/dx$ curve obtained from each division exhibits good statistical
convergence in center of its domain and poorer statistical convergence
at the margins.  Within statistical uncertainty, the $dE_j/dx$ curves
are consistent with
the potential used in the simulation and with each other.  
When the 20 $dE_j/dx$ curves obtained from the 20 divisions are combined 
using Eq.~\ref{eq:DESArewrite} good agreement 
is found between the reconstructed $dE/dx$ shown in Fig.~\ref{fig:SdEdx}(d)
and the energy surface used in the simulation.

When the same analysis is applied to system II the DESA method
provides a visual indication that the 
dynamics of the system are not described by an energy which 
can be expressed as a function of a single reaction coordinate $x$.
In Fig.~\ref{fig:Strajectory}(e)-(g) $dE_j/dx$ estimates
from three divisions of the trajectory are shown.  They are
compared with derivative of the system energy with respect to $x$, 
assuming that the 
system remains in equilibrium with respect to $y$ (solid curve),
assuming that the system is confined to negative $y$ (dashed curve)
and assuming that the system is confined to positive $y$ (short dashed
curve).
The estimate of $dE_j/dx$ obtained from division 5 conforms to the 
potential for negative $y$ while the estimate from divisions 10 
and 15 conform to the 
potential for positive $y$.  
The reconstructed $dE/dx$ curve shown in Fig.~\ref{fig:SdEdx}(h) 
gives a smooth curve despite the fact that it
is obtained from averaging of inconsistent $dE_j/dx$ functions.

We next apply the DESA diagnostics introduced in Section~\ref{sec:diagnostics}.
In Fig.~\ref{fig:Schi2} the reduced $\chi^2$ test defined
in Eq.~\ref{eq:chi2} is applied to both simulated systems.
The purpose of this test is to determine if the values of $dE_j/dx$ obtained
from the various divisions of the data are consistent with each other,
taking into account the statistical uncertainties of the various estimates.
Fig.~\ref{fig:Schi2}(a) shows that for system I 
the reduced $\chi2$ is of order 1 over the full range
of $x$.  This indicates that the $dE_j/dx$ functions obtained for the different
constraint origins are mutually consistent.
However, Fig.~\ref{fig:Schi2}(b) shows that for system II
the value of the reduced $\chi^2$ function
increases to 15 in the vicinity of the apparent transition state.  This confirms
our observation that in Fig.~\ref{fig:SdEdx}(e)-(g) 
the $dE_j/dx$ functions deviate from each other 
by an amount exceeding the statistical
uncertainty.  This alerts us that data obtained from different biasing potentials
are not consistent and $dE/dx$ is not a well defined function of $x$,
despite the fact that the curve obtained
is smooth and appears plausible.  

Next we consider the diagnostic criteria defined in Eq.~\ref{eq:diag1},
in which we subtract 
the raw energy surfaces obtained from data taken with 
different biasing potentials.  
In Fig.~\ref{fig:Sslope}(a), Eq.~\ref{eq:diag1} is
evaluated for representative divisions of the data from system I.  Linear 
curves are found with slopes that are consistent with the constraint stiffness
used in the simulation.  
In Fig.~\ref{fig:Sslope}(b) the same measure is applied
to representative divisions from system II.  
Inconsistent slopes, or non-linear curves are observed.
This indicates that the intrinsic energy surface of the system
failed to cancel when the energy surfaces of different divisions were
subtracted.
As in the case of Fig.~\ref{fig:Schi2}, it is evident that the data
produced by the simulation of system II are not self-consistent.

The final question we can address is whether DESA or WHAM are more accurate in
determining the relative energy of the initial and final states for the two systems.
In Fig.~\ref{fig:SEnergy} we compare the energy surfaces obtained by
direct integration of the DESA $dE/dx$ curve, and 
using WHAM.
In the case of the well-behaved system I (Fig.~\ref{fig:SEnergy}(a)), 
both DESA and WHAM produce energy curves
which match the potential used in the simulation.  
In the case of system II (Fig.~\ref{fig:SEnergy}(b)) we find that 
DESA and WHAM produce energy curves which are
effectively identical.
Both curves fail to agree with the actual
energy difference between the initial and final state.
In this example, the energy surface was known \emph{a priori}
making direct comparison possible.  
However, the diagnostic criteria illustrated in Fig.~\ref{fig:Schi2}
and \ref{fig:Sslope} alerted us to problems in the reconstruction of the energy
surface and did not require
knowledge of the correct energy surface.

\section{Application of DESA to experimental data}
\label{sec:experiment}

Here we apply DESA and WHAM 
to a single molecule experiment in which a DNA hairpin is unfolded
under a harmonic constraint applied by an optical trap.
The hairpin has sequence
C\-C\-G\-C\-G\-A\-G\-T\-T\-G\-A\-T\-T\-C\-G\-C\-C\-A\-T\-A\-C\-A\-C\-C\-T\-G\-C\-T\-A\-A\-T\-C\-C\-C\-G\-G\-T\-C\-G\-C\-T\-T\-T\-T\-G\-C\-G\-A\-C\-C\-G\-G\-G\-A\-T\-T\-A\-G\-C\-A\-G\-G\-T\-G\-T\-A\-T\-G\-G\-C\-G\-A\-A\-T\-C\-A\-A\-C\-T\-C\-G\-C\-G\-G,
which folds into a 
40 base-pair stem with a
4-T loop.  
The hairpin is connected to the boundary of the sample chamber on one side
and to a polystyrene micro-sphere on the other via biotin and digoxigenin 
tagged double-stranded
DNA linkers.
This creates a single-molecule tether which anchors the micro-sphere to the
surface.  When the optical trap is held at constant
position and intensity, the combination of the restoring force 
imposed on the micro-sphere by
the optical trap and the elasticity of the handles
produce a harmonic constraint acting on the hairpin with
$\alpha \approx 200~\mathrm{pN}/\mathrm{\mu m}$.  
The position of the sample chamber relative to the trapping beam is controlled
by a piezoelectric positioning stage with nanometer resolution, 
and the origin of the constraint is controlled by varying
the position of the sample chamber with respect to the trap center.  
The optical trap measures the instantaneous position of the micro-sphere
and the instantaneous force applied to the tether as the constraint
origin is swept.  By determining 
the distance between the micro-sphere and the sample chamber boundary
and subtracting off the instantaneous extension of the double-stranded
DNA handles (estimated using a worm-like chain model of DNA elasticity) the
extension of the hairpin itself is determined\cite{marko:1995}.
The apparatus and experimental procedure has been
described elsewhere\cite{demessieres:2011}.
The measured time series comprised of approximately $10^5$ samples
is shown for unfolding of the hairpin in Fig.~\ref{fig:timeseries}(a),
and for folding of the hairpin in Fig.~\ref{fig:timeseries}(b). 

As in the case of the simulated data, the 
record of extension vs.\ time of the experimental system is
divided into 20 equal time intervals and the histogram of position
is calculated for each division.  
Calibration data is used to calculate
the mean stiffness and origin of the constraint
for each of the 20 divisions\cite{demessieres:2011}.
Since the constraint origin
moves continuously as data is acquired, 
it is not a constant within each interval.
However, the deviation of the constraint origin from the
mean value does not exceed
$\sim$1~nm in the course of an interval, which implies an error
in the constraint force of less than $\sim$0.2~pN.  The
resulting error in the reconstruction of $dE/dx$ is negligible.

In Fig.~\ref{fig:dEdxD} the $dE_j/dx$ functions calculated
from representative divisions of the data in Fig.~\ref{fig:timeseries}(a)
are shown in panels (b) through (f) and the $dE/dx$ function
obtained by averaging all 20 divisions is shown in panel (a).
Just as in Fig.~\ref{fig:SdEdx}(a)-(c), the individual $dE_j/dx$
estimates in Fig.~\ref{fig:dEdxD} are consistent with each other and with
the average $dE/dx$ function within statistical uncertainty.
This justifies the assumption that the experimental system
continues to explore the same energy landscape as the constraint origin
moves.

As in the simulated system, the umbrella sampling method requires us to assume
that the biasing potential is time independent and that the 
system remains in thermodynamic equilibrium as data is collected.
Since the constraint origin moves continuously as data is collected
this condition is not formally satisfied, and we must
insure that the movement of the constraint is sufficiently slow that the
system remains in equilibrium to good approximation.  
The most convincing evidence that this condition is satisfied
is that identical energy landscapes are obtained for folding and
unfolding of the hairpin, for which the constraint origin moves in opposite
directions.  The energy landscapes obtained from data
in Figs.~\ref{fig:timeseries}(a) and \ref{fig:timeseries}(b) are compared
in Fig.~\ref{fig:reverse}.  No significant difference is found between
landscapes obtained for folding and unfolding of the hairpin.

In contrast with the simulations, the effective energy of the hairpin is not known
\emph{a priori} so the diagnostic criteria introduced in 
Section~\ref{sec:diagnostics} are of critical importance in establishing
the validity of the energy landscape reconstruction.
In Fig.~\ref{fig:Ediagnostic} we apply the two diagnostic criteria
defined in Section~\ref{sec:diagnostics} to the experimental data set.
In Fig.~\ref{fig:Ediagnostic}(a) the $\chi^2$ measure is plotted on a logarithmic
scale.  Note that for the central region of the reaction coordinate, corresponding
to the transition state, the value of $\chi^2$ is of order unity, which indicates 
that the $dE_j/dx$ curves obtained in the transition state region 
are consistent within statistical uncertainty.  This confirms that
a well-defined
energy function is being measured.
At the extremes (near extension 0~$\mu$m and 0.05~$\mu$m) the value of 
$\chi^2$ is larger, indicating that the $dE_j/dx$ curves are inconsistent
at large and small extension.  

The larger $\chi^2$ values are found in regimes of extension where the
hairpin is either fully open or fully folded.   When the constraint is positioned
to stabilize the hairpin in the fully open or fully closed conformation, 
the conformational dynamics of the hairpin itself are minimal and the fluctuations in
the measured extension are mainly due to thermal fluctuation in the extension
of the double-stranded DNA handles.  
At small extensions, the problem is exacerbated by the fact that
the average force is low, resulting in lower 
effective stiffness of the handles and
increased fluctuations.  These measurement errors blur the sharp
cutoff that would otherwise appear in the probability density of 
extension as the
hairpin approaches the fully-open or fully-folded state and similarly
blur the energy function.  The $\chi^2$ function alerts us to the fact that 
the energy surface is accurately measured in the transition state region, but
is affected by systematic errors near the fully folded or fully unfolded state.

We also apply the criterion based on Eq.~\ref{eq:diag1} and show the results
in Fig.~\ref{fig:Ediagnostic}(b).
The fact that linear curves are obtained when the unbiased energies are
subtracted indicate that the intrinsic energy of the hairpin cancels, as expected,
and that the effective biasing potential has the expected parabolic shape.
This is confirmation that the optical trapping apparatus is applying an
accurate biasing potential to the hairpin.
Based on Fig.~\ref{fig:Ediagnostic} we conclude that the energy of the hairpin
is a well-defined function of extension and that DESA has produced an
accurate measurement of the transition state region.

In order to verify the DESA result, the data shown in Fig.~\ref{fig:timeseries} was
also analyzed using WHAM,
as defined by Eqs ~\ref{eq:WHAM1} and \ref{eq:WHAM2}.
In Fig.~\ref{fig:DESAWHAM}(a) the energy surface 
obtained by WHAM is plotted
along with the energy surface obtained from integration of the
$dE/dx$ curve shown in Fig.~\ref{fig:dEdxD}.  The DESA and WHAM curves are
indistinguishable.  The overall slope of the energy landscape
is reproduced, as well as the ripples that arise from the sequence
dependence of the DNA hybridization energy.  The sequence
dependence is more apparent in the plot of $dE/dx$, which 
is shown in Fig.~\ref{fig:DESAWHAM}(b).  As in the case
of the energy, results obtained from DESA and WHAM 
are indistinguishable.

\section{Conclusions}

There are systems, such as pseudoknots, G-quadruplex DNA and others, which
exhibit large irreversible steps when disrupted
in single molecule experiments\cite{green:2008,yu:2009,demessieres:2012}.  
In such cases, the techniques
described here would not be suitable for reconstructing the global energy landscape.  
The main obstacle is that it is impossible to apply a biasing potential which will
stabilize the system in the transition state or states.  Both WHAM and DESA
require that the histograms of the reaction coordinate obtained with different biasing
potentials have substantial overlap.
Nonequilibrium analysis
methods have been developed which can determine the energy surface
from data taken far from equilibrium\cite{jarzynski:1997,crooks:2000,hummer:2001,hummer:2010,evans:2001,dudko:2008}.  
These methods typically require a great deal of experimental data, 
since they involve measuring the dependence of the disruption
force on force loading rate or averaging
many trajectories with weights determined by the external work
performed.

In cases where biasing potentials can be used to stabilize a system
along the reaction coordinate, DESA is an alternative to WHAM.  
We have shown that DESA and WHAM produce indistinguishable results when applied
to simulated and experimental data.  
However, DESA has the advantage of being computationally simple compared
with WHAM, which requires
the self-consistent solution of a system of nonlinear equations (Eq.~\ref{eq:WHAM2}).  
Another advantage of DESA is that the construction of $dE/dx$
provides direct visual cues which can be used to confirm that 
the different biasing potentials are sampling the same energy surface
(see Fig.~\ref{fig:dEdxD}).
In addition, the two diagnostic criteria defined in Section~\ref{sec:diagnostics}
provide quantitative measures of the quality of the energy surface
measurement.
The signatures of an ill-defined energy surface are demonstrated
in the analysis of data generated by simulation system II.
Finally, using the DESA diagnostics, we show that it is possible
to apply a precisely controlled biasing potential in an experimental
system and obtain highly accurate information about the shape of the
energy surface for folding and unfolding.

This work was supported by the Maryland Technology Development Corporation.

%


\end{document}